\documentclass{amsart}

\makeatletter

\newtheorem{thm}{Theorem}[section]
\def\statetheorem{\@ifnextchar[{\@statetheorem}{\nr@statetheorem}}
\long\def\@statetheorem[#1]#2{\begin{thm}\label{#1}#2\end{thm}}
\long\def\nr@statetheorem#1{\begin{thm}#1\end{thm}}
\def\statetheorempf{\@ifnextchar[{\@statetheorempf}{\nr@statetheorempf}}
\long\def\@statetheorempf[#1]#2{\begin{thm}\label{#1}#2\end{thm}\proof}
\long\def\nr@statetheorempf#1{\begin{thm}#1\end{thm}\proof}

\newtheorem{lmma}{Lemma}[section]
\def\statelemma{\@ifnextchar[{\@statelemma}{\nr@statelemma}}
\long\def\@statelemma[#1]#2{\begin{lmma}\label{#1}#2\end{lmma}}
\long\def\nr@statelemma#1{\begin{lmma}#1\end{lmma}}
\def\statelemmapf{\@ifnextchar[{\@statelemmapf}{\nr@statelemmapf}}
\long\def\@statelemmapf[#1]#2{\begin{lmma}\label{#1}#2\end{lmma}\proof}
\long\def\nr@statelemmapf#1{\begin{lmma}#1\end{lmma}\proof}

\newtheorem{crlry}{Corollary}[section]
\def\statecorollary{\@ifnextchar[{\@statecorollary}{\nr@statecorollary}}
\long\def\@statecorollary[#1]#2{\begin{crlry}\label{#1}#2\end{crlry}}
\long\def\nr@statecorollary#1{\begin{crlry}#1\end{crlry}}
\def\statecorollarypf{\@ifnextchar[{\@statecorollarypf}{\nr@statecorollarypf}}
\long\def\@statecorollarypf[#1]#2{\begin{crlry}\label{#1}#2\end{crlry}\proof}
\long\def\nr@statecorollarypf#1{\begin{crlry}#1\end{crlry}\proof}

\def\ord{\textup{ord}\,}

\def\C{\mathbb{C}}
\def\N{\mathbb{N}}
\def\bibref[#1]{\cite{#1}}

\let\real@bibitem\bibitem
\def\bibitem[#1]{\real@bibitem{#1}}

\makeatother

\title{KP Solitons are Bispectral}

\author{Alex Kasman}
\address{Mathematical Sciences Research Institute\\Berkeley, CA 94720}

\def\genT{{\hat T}}

\def\hatLambda{{\hat\Lambda}}

\def\D{{\Bbb D}}

\def\TODOs{{\Bbb T}}
\def\L{{\mathcal L}}
\def\Trans{{\textbf{S}}}
\begin{document}



\begin{abstract}{It is by now well known that the wave functions of rational
solutions to the KP hierarchy which can be achieved as limits of the
pure $n$-soliton solutions satisfy an eigenvalue equation for ordinary
differential operators in the spectral parameter.  This property is
known as ``bispectrality'' and has proved to be both interesting and
useful.  In a recent preprint (math-ph/9806001) evidence was presented
to support the conjecture that all KP solitons (including their
rational degenerations) are bispectral if one also allows translation
operators in the spectral parameter.  In this note, the conjecture is
verified, and thus it is shown that all KP solitons have
a form of bispectrality.  The potential significance of this result to
the duality of the classical Ruijsenaars and Sutherland particle
systems is briefly discussed.}\end{abstract}
\maketitle

\section{Introduction}
\subsection{The KP Hierarchy and Bispectrality}
Let $\D$ be the vector space spanned over $\C$ by the set
$$
\{\Delta(\lambda,n)\mid \lambda\in\C, n\in\N \}
$$
whose elements differentiate and evaluate functions of the variable
$z$:
$$
\Delta(\lambda,n)[f(z)]:=f^{(n)}(\lambda).
$$
The elements of $\D$ are thus \emph{finitely supported
  distributions} on appropriate spaces of functions in $z$.  For lack
of a better term, we will continue to call them distributions even
though their main use in this paper will be their application to
functions of two variables.    (Such
distributions were called ``conditions'' in \bibref[W] since a
KP wave function was specified by requiring that it be in their
kernel.)  Note that if $c\in\D$ and $f(x,z)$
is sufficiently differentiable in $z$ on the support of $c$, then
$\hat f(x)=c[f(x,z)]$ is a function of $x$ alone.  Furthermore,
note that one may ``compose'' a distribution with a function of $z$,
i.e. given $c\in\D$ and $f(z)$ (sufficiently differentiable on the
support of $c$) then there exists a $c':=c\circ f\in\D$ such
that
$$
c'(g(z))=c(f(z)g(z))\qquad\forall g.
$$

The subspaces of $\D$ can be used to generate solutions to the KP
hierarchy \bibref[SW] in the following way.  Let $C\subset\D$ be an
$n$ dimensional subspace with basis $\{c_1,\ldots,c_n\}$.  Then, if
$K=K_C$ is the unique, monic ordinary differential operator in $x$ of
order $n$ having the functions $c_i(e^{xz})$ in its kernel (see
\eqref{eqn:wrformK}) we define $\L_C=K \frac{\partial}{\partial x} K^{-1}$ and
$\psi_C=\frac{1}{z^n}K e^{xz}$.  The connection to integrable systems
comes from the fact that adding dependence to $C$ on a sequence of
variables $t_j$ ($j=1,2,\ldots$) by letting $C(t_j)$ be the space with
basis $$\{c_1\circ e^{\sum -t_jz^j},c_2\circ e^{\sum -t_jz^j},\ldots,
c_n\circ e^{\sum -t_jz^j}\}$$ it follows that the ``time dependent''
pseudo-differential operator $\L=\L(t_j)$ satisfies the equations of
the KP hierarchy \bibref[BHYsato,thesis,cmbis,W] $$
\frac{\partial}{\partial t_j}\L=[(\L^j)_+,\L].
$$
The {\it wave function\/}
$\psi_C(x,z)$ generates the corresponding subspace of the infinite
dimensional grassmannian $Gr$ \bibref[SW] which parametrizes KP solutions and
thus it is not difficult to see that this construction produces
precisely those solutions associated to the subgrassmannian
$Gr_1\subset Gr$ \bibref[SW,W].

Moreover, the ring $A_C=\{p\in\C[z]|c_i\circ p\in C\ 1\leq i \leq n\}$
is necessarily non-trivial (i.e.\ contains non-constant polynomials) and the operator $L_p=p(\L)$ is an {\it
ordinary\/} differential operator for every $p\in A_C$ and satisfies
\begin{equation}
L_p\psi_C(x,z)=p(z) \psi_C(x,z).\label{eigenx}
\end{equation}
The subject of this paper is the existence of {\it additional\/} eigenvalue
equations satisfied by $\psi_C(x,z)$.  In particular, we wish to
consider the question of whether there exists an
operator $\hatLambda $ acting on functions of the variable $z$ such that
\begin{equation}
\hatLambda \psi_C(x,z)=\pi(x)\psi_C(x,z)\label{eigenz}
\end{equation}
where $\pi(x)$ is a non-constant function of $x$.
For example, the following theorem is due to G. Wilson in \bibref[W]:
\statetheorem[Th:wilson]{In addition to \eqref{eigenx} the wave function $\psi_C(x,z)$ is also an
eigenfunction for a ring of ordinary differential operators in $z$
with eigenvalues depending polynomially on $x$ if and only if $C$ has a
basis of distributions each of which is supported only at one point.}
In other words, for this special class of KP solutions for which the
coefficients of $\L$ are rational functions of $x$, the wave
function $\psi_C$ satisfies an additional eigenvalue equation of the form
\eqref{eigenz}
where $\hatLambda$ is an ordinary differential operator in $z$ and
$\pi(x)$ a non-constant polynomial in $x$.\footnote{Moreover, he
demonstrated that up to conjugation or change of variables, the
operators $L_p$ found in this way are the only bispectral operators
which commute with differential operators of relatively prime order,
but this fact will not play an important role in the present note.}
Together \eqref{eigenx} and \eqref{eigenz} are an example of {\it
bispectrality\/} \bibref[DG,G].  The bispectral property is already
known to be connected to other questions of physical significance such
as the time-band limiting problem in tomography \bibref[Gr1], Huygens'
principle of wave propagation \bibref[Yuri], quantum integrability
\bibref[HK,Vbis] and, especially in the case described above, the self
duality of the Calogero-Moser particle system \bibref[cmbis,W,W2].

It is known that the only subspaces $C$ for which the corresponding
wave function satisfies \eqref{eigenx} and \eqref{eigenz} with $L_p$ and
$\hatLambda $ ordinary differential operators in $x$ and $z$ respectively
are those described in Theorem~\ref{Th:wilson}.  However, suppose we
allow $\hatLambda$ to involve not only differentiation and multiplication
in $z$ but also {\it translation\/} in $z$ and  call this more {\it
general\/} situation t-bispectrality.\footnote{It should be noted that
the term ``bispectrality'' already applies to more general situations
than simply differential operators \bibref[G], but in the case of the
KP hierarchy I believe only differential bispectrality has thus far
been considered.}  It will be shown below that there are more KP
solutions which are bispectral in this sense.  In particular, it will
be shown that the KP solution associated to \textit{any} subspace $C$
shares its eigenfunction with a ring of translational-differential
operators in the spectral parameter.

\subsection{Notation}


Using the shorthand notation 
$\partial=\frac{\partial}{\partial x}$ any ordinary differential
operator in $x$ can be
written as
$$
L=\sum_{i=0}^N f_i(x) \partial^i\qquad (N\in\N).
$$
We say that a function of the form
$$
f(x)=\sum_{i=1}^n p_i(x)e^{\lambda_i x}\qquad \lambda_i\in\C,\
p_i\in\C[x]
$$
is a \textit{polynomial-exponential function} and that the quotient of
two such functions is \textit{rational-exponential}.  
This note will be especially concerned with the ring of differential
operators with rational-exponential coefficients and especially with
the subring having polynomial-exponential coefficients.
Similarly, we will write
$\partial_z=\frac{\partial}{\partial z}$ but will need to consider
only differential operators in $z$ with rational coefficients.

For any $\lambda\in\C$ let $\Trans_{\lambda}=e^{\lambda\partial_z}$ be the
translational operator acting on functions of $z$ as
$$
\Trans_{\lambda}(f(z))=f(z+\lambda).
$$
Then consider the ring of translational-differential operators
$\TODOs$ generated by these translational operators and ordinary
differential operators in $z$.  Any translational-differential operator $\hat
T\in\TODOs$  can be
written as
$$
\genT =\sum_{i=1}^N p_i(z,\partial_z) \Trans_{\lambda_i}
$$
where $p_i$ are ordinary differential operators in $z$ with rational
coefficients  and $N\in\N$.  Note that the ring of ordinary
differential operators in $z$ with rational coefficients is
simply the subring of $\TODOs$ of all elements which can be written as
$p \Trans_{0}$ for a differential operator $p$.

\section{Translational Bispectrality of $\C[\partial]$}

It has been frequently observed that the ring
$\mathcal{A}=\C[\partial]$ of constant coefficient differential
operators in $x$ is \textit{bispectral} since it has the eigenfunction
$e^{xz}$ which it shares with the ring of constant coefficient
differential operators in $z$.  Here, however, we will consider a more
general form of bispectrality for the ring $\mathcal{A}$.  

Let $\mathcal{A}'\subset\TODOs$ be the ring of constant
coefficicient \textit{translational}-differential operators.  Note
that for any element $\genT\in\mathcal{A}'$ of the form
$$\genT =\sum_{i=1}^N p_i(\partial_z) \Trans_{\lambda_i}$$
one has simply that
$$
\genT[e^{xz}]=\left(\sum_{i=1}^N p_i(x)e^{\lambda_i x}\right)e^{xz}.$$
In particular, $e^{xz}$ is an eigenfunction for the operator with an
eigenvalue which is a polynomial-exponential function of $x$.
Consequently, the rings $\mathcal{A}$ and $\mathcal{A}'$ are
both bispectral, sharing the common eigenfunction $e^{xz}$.

Let $\mathcal{R}$ be the ring of differential operators in $x$ with
polynomial-exponential coefficients and $\mathcal{R'}$ be the ring of
translational-differential operators in $z$ with rational
coefficients.  Note that $\mathcal{R}$ is generated by $\mathcal{A}$
and the eigenvalues of the operators in $\mathcal{A'}$ while
$\mathcal{R}'$ is generated by $\mathcal{A}'$ and the eigenvalues of
the elements of $\mathcal{A}$.  It then follows \bibref[BHYpla] (see also
\bibref[KR]) that the map $b:\mathcal{R}\to\mathcal{R}'$ defined by
the relationship
$$
L[e^{xz}]=b(L)[e^{xz}]\qquad \forall L\in\mathcal{R}
$$
is an anti-isomorphism of the two rings.

\section{Translational Bispectrality of KP Solitons}

Let us say that a finite dimensional
subspace $C\subset\D$ is {\it t-bispectral\/} if there exists a
translational-differential operator  $\hatLambda \in \TODOs$ satisfying
equation \eqref{eigenz} for the corresponding KP wave function
$\psi_C(x,z)$.
By Theorem~\ref{Th:wilson} and the fact that the ring of rational coefficient
ordinary differential operators in $z$ is contained in $\TODOs$, we
know that $C$ is t-bispectral\footnote{...and also bispectral
in the sense of \bibref[W].} if it has a basis of point supported
distributions.  Here we will show that, in fact, all subspaces
$C\subset\D$ are $t$-bispectral.

An important object in much of the literature on integrable systems is
the ``tau function''.  The tau function of the KP solution associated
to $C$ can be written easily in terms of the basis $\{c_i\}$.  In
particular, define (cf.\ \bibref[W])
$$
\tau_C(x)=\textup{Wr}\left(c_1(e^{xz}),c_2(e^{xz}),\ldots,c_n(e^{xz})\right)
$$
to be the Wronskian determinant of the functions $c_i(e^{xz})$.
Similarly, there is a Wronskian formula for the coefficients of the
operator $K_C$ since its action on an arbitrary function $f(x)$ is
given as:
\begin{equation}
K_C(f(x))=\frac{1}{\tau_C(x)} \textup{Wr}\left(c_1(e^{xz}),c_2(e^{xz}),\ldots,c_n(e^{xz}),f(x)\right).\label{eqn:wrformK}
\end{equation}
Then the coefficients of the differential operator $\bar
K_C:=\tau_C(x)K_C(x,\partial)$ are all polynomials-exponential
functions.

\begin{lemma}[lem:factor]
{For any $C\subset\D$ there exists a constant coefficient operator
$L_0\in\mathcal{A}$ which factors as
$$
L_0=\bar Q\circ\frac{1}{\pi(x)}\circ \bar K_C
$$
where $\bar Q,\bar K_C\in\mathcal{R}$ and $\pi(x)\in\mathcal{R}$ is a
polynomial-exponential function.}
Let $\lambda_i\in\C$ ($1\leq i \leq N$) be the support of the
distributions in $C$ and $m_i$ be the highest derivative taken at
$\lambda_i$ by any element of $C$.  Then the polynomial
\begin{equation}
q_C(z):=(z-\lambda_i)^{m_i+1}\label{eqn:q}
\end{equation}
 has the property that $c\circ q_C\equiv0$
for any $c\in C$.  Let $L_0:=q_C(\partial)$ and consider
$L_0[c(e^{xz})]$ for any element $c\in C$.  Since $L_0$ is an operator
in $x$ alone, it commutes with $c$ and we have
$$
L_0[c(e^{xz})]=c\left(L_0[e^{xz}]\right)=c\left(q(z)e^{xz}\right)=c\circ
q(e^{xz}=0.
$$
So, by the definition of $K_C$, we see that $L_0$ annihilates the
kernel of $K_C$ and thus has a right factor of $K_C$.  This gives a
factorization of the form $L_0=Q\circ K_C$ with $Q$ having
rational-exponential coefficients.  Then, by choosing a
polynomial-exponential function $g(x)$ so that $\bar Q:=Q\circ
g(x)\in\mathcal{R}$ we find the desired factorization with
$\pi(x)=g(x)\tau_C(x)$. 
\end{lemma}

Given this factorization, the t-bispectrality of all $C$'s now follows
from Theorem 4.2 in \bibref[BHYpla]. 
\begin{theorem}[thm:main]
{For any subspace $C\subset\D$ the
translational-differential operator $\hatLambda\in\TODOs$ 
defined by $$\hatLambda:=z^{-n}\circ b(\bar K_C)\circ b(\bar Q)\circ \frac{z^n}{q_C(z)}
$$
satisfies the equation
$$
\hatLambda[\psi_C(x,z)]=\pi(x)\psi_C(x,z)
$$
with $\pi(x)$ the polynomial-exponential function of $x$ from the
factorization above.}
Formally introducing inverses \bibref[BHYpla], we determine from
Lemma~\ref{lem:factor} that
$$
\pi(x)=\bar K_C \circ L_0^{-1} \circ \bar Q
$$
and hence (by applying the anti-involution $b$ to this equation)
$$
b(\pi(x))=b(\bar Q)\circ \frac{1}{q_C(z)}\circ b(\bar K_C).
$$
Then 
\begin{eqnarray*}
\hatLambda[\psi_C(x,z)] &=& 
z^{-n}\circ b(\bar K_C)\circ b(\bar Q)\circ
\frac{z^n}{q_C(z)}[\frac{1}{z^n \tau_C(x)} \bar K_C e^{xz}]\\
 &=& \frac{z^{-n}}{\tau_C(x)} \circ b(\bar K_C)\circ b(\bar Q)\circ
\frac{1}{q_C(z)}[\bar K_C e^{xz}]\\
 &=& \frac{z^{-n}}{\tau_C(x)} \circ b(\bar K_C)[\pi(x)e^{xz}]\\
 &=& \frac{z^{-n}\pi(x)}{\tau_C(x)} \circ \bar K_C[e^{xz}]\\
 &=& \pi(x)\psi_C(x,z)
\end{eqnarray*}
\end{theorem}

\section{Examples}

If we choose $C$ to be the two dimensional space spanned by $c_1=\Delta(1,0)$ and $c_2=\Delta(1,1)$ (a ``two-particle''
Calogero-Moser type solution) then
$$
\psi_C(x,z)=(1+\frac{2+x-(2x+x^2)z}{x^2z^2})e^{xz}.
$$
In this case the translational-differential operator $\hatLambda$
given by Theorem~\ref{thm:main} is simply an ordinary differential
operator.  For instance,
$$
\hatLambda =\partial_z^3+\frac{3}{z-z^2}\partial_z^2-\frac{6z^2-12z+3}{z^3(z-1)^2}\partial
+ \frac{12z-6}{z^2(z-1)^2}$$
which satisfies $\hatLambda \psi_C(x,z)=x^3\psi_C(x,z)$ (as we would
expect from earlier results on bispectrality.)

However, if we had chosen instead $c_1=\Delta(0,1)+\Delta(0,-1)$ and
$c_2=\Delta(0,2)+\Delta(0,0)$ we would instead have 
the case of a 2-soliton solution with
$$
\psi_C(x,z)=(1-\frac{6+(3z-2)e^{2x}+2z-ze^{-2x}}{(e^x+e^{-x})^2z^2})e^{xz}.
$$
One finds from the procedure given in the theorem that 
\begin{eqnarray*}
\hatLambda   &=&  z^{-2}\circ\left( 
\left( 20z + 11{z^2} - 8{z^3} + 
     {z^4} \right)\Trans_{-3}
  + 
   \left( 60 - 68z - {z^2} + 8{z^3} + {z^4} \right) \Trans_{5}
  \right.\\
&& + 
  \left( -36 + 24z + 16{z^2} - 16{z^3} + 
     4{z^4} \right)\Trans_{-1}
 + 
   \left( -44 - 88z - 8{z^2} + 16{z^3} + 4{z^4} \right) \Trans_{3}\\
&&
\left.\left( -12 - 16z - 2{z^2} + 
     6{z^4} \right)\Trans_{1}
\right)\circ\frac{z^n}{z^4-2z^3-z^2+2z}
\end{eqnarray*}
satisfies $\hatLambda \psi_C(x,z)=e^{-3x}(1+e^{2x})^4\psi_C(x,z)$.

\section{Conclusions}

In addition to being a generalization of the results of \bibref[DG,W]
on bispectral ordinary differential operators, the present note may be
seen as a generalization of \bibref[Reach] in which wave functions of
$n$-soliton solutions of the KdV equation are shown to satisfy
difference equations in the spectral parameter. 
The idea that KP solitons might be translationally bispectral was
proposed in \bibref[KPsol1].

As in \bibref[DG,W], the equations \eqref{eigenx} and \eqref{eigenz}
lead to the well known ``ad'' relations associated to bispectral
pairs.
That is, defining the ordinary differential operator $A_m$ in $x$ and
the translational-differential operator $\hat A_m$ in $z$ by 
$$A_m=\textup{ad}_{L_p}^m(\pi(x))
\qquad
\hat A_m=(-1)^m\textup{ad}_{p(z)}^m(\hatLambda)
$$
one finds that $A_m\psi_C(x,z)=\hat A_m\psi_C(x,z)$.  Similarly, if
$$B_m=\textup{ad}_{\pi(x)}^m(L_p)
\qquad
\hat B_m=(-1)^m\textup{ad}_{\hatLambda}^m(p(z))
$$
then $B_m\psi_C(x,z)=\hat B_m\psi_C(x,z)$.  Since the order of the
operator $B_m$ is at least one less than the order of the operator
$B_{m-1}$, the familiar result that $B_m\equiv0$ and $\hat B_m\equiv0$
for $m>\ord L_p$ holds, which is clearly a strong restriction on
the operator $\hatLambda$.  However, unlike the case of bispectral
ordinary differential operators, one cannot conclude that $A_m\equiv0$
for sufficiently large $m$ since the order of $\hat A_m$ may not be
reduced by increasing $m$.

The function $g(x)$ in Lemma~\ref{lem:factor} is not unique.  Thus it
might be more reasonable to write $\hatLambda_g$ as the
translational-differential operator having $\psi_C$ as an
eigenfunction.  It then follows that the set of these operators for
all choices of $g$ forms a commutative ring.

The bispectrality of the rational KP solutions \bibref[W] has been
shown to have a dynamical significance.  In particular, it was shown
that the \textit{bispectral involution} is the linearizing map for the
classical Calogero-Moser particle system \bibref[cmbis,W,W2].  Moreover, other
bispectral KP solutions have been found to have similar properties
\bibref[cmbis2,Roth].  This would seem to indicate that it is likely
that the bispectrality of KP solitons also has a dynamical
significance, as a map between the classical Ruijsenaars and Sutherland
systems.  In fact, such a bispectral relationship between the
\textit{quantum} versions of these systems has been recently found in
\bibref[chal].  The dynamical significance of these results will be
considered in a separate paper.


\end{document}